\pdfoutput=1
\documentclass[manuscript,bookmarksnumbered,unicode,colorlinks=true,linkcolor=blue, hyperfootnotes=true]{acmart}

\makeatletter                   
\def\mdseries@tt{m}             
\makeatother                    
\usepackage[plain]{fancyref}
\usepackage[draft=true]{minted} 
\usepackage{color}
\usepackage{breakurl}           

\usepackage{caption}
\usepackage{subcaption}

\begin{document}

\title{Expressive Communication: A Common Framework for Evaluating Developments in Generative Models and Steering Interfaces}

\author{Ryan Louie$^{*}$}
\affiliation{%
  \institution{Northwestern University}
  \city{Evanston}
  \country{IL}
}
\email{ryanlouie@u.northwestern.edu}

\author{Jesse Engel}
\affiliation{%
  \institution{Google Research}
  \city{Mountain View, CA}
  \country{USA}
  }
\email{jesseengel@google.com}

\author{Anna Huang}
\affiliation{%
  \institution{Google Research}
  \city{Montreal, Quebec}
  \country{Canada}
  }
\email{annahuang@google.com}

\renewcommand{\shortauthors}{Louie, et al.}
\renewcommand{\shorttitle}{Expressive Communication: A Framework for Evaluating Generative Models and Interfaces}

\begin{abstract}

There is an increasing interest from ML and HCI communities in empowering creators with better generative models and more intuitive interfaces with which to control them.
In music, ML researchers have focused on training models capable of generating pieces with increasing long-range structure and musical coherence, while HCI researchers have separately focused on designing steering interfaces that support user control and ownership.
In this study, we investigate through a common framework how developments in both models and user interfaces are important for empowering co-creation where the goal is to create music that communicates particular imagery or ideas (e.g., as is common for other purposeful tasks in music creation like establishing mood or creating accompanying music for another media). 
Our study is distinguished in that it measures communication through both composer's self-reported experiences, and how listeners evaluate this communication through the music. In an evaluation study with 26 composers creating 100+ pieces of music and listeners providing 1000+ head-to-head comparisons, we find that more expressive models and more steerable interfaces are important and complementary ways to make a difference in composers communicating through music and supporting their creative empowerment.

\end{abstract}


\maketitle

\renewcommand*{\thefootnote}{\fnsymbol{footnote}}
\setcounter{footnote}{1}
\footnotetext{This work was performed during the first author's summer internship at Google Research.}
\renewcommand*{\thefootnote}{\arabic{footnote}}
\setcounter{footnote}{0}

%

\section{Introduction}

There is an increasing interest from machine learning (ML) and human computer interaction (HCI) communities in empowering creators with better generative models and more intuitive interfaces with which to control them.
In the domain of music, ML researchers have focused on training models capable of generating pieces with increasing long-range structure and musical coherence~\cite{huang2018music,payne2019muse}, while HCI researchers have separately focused on designing better steering interfaces that support user control and ownership through overcoming AI-induced information overload and non-deterministic model outputs~\cite{louie2020noviceai}. 

While the ML and HCI communities have similar aspirations for generative modeling, interdisciplinary collaborations have been limited by a lack of common frameworks to evaluate progress.
Many ML researchers desire their models to be useful for creators, but most evaluations of these generative models stop short of directly measuring and optimizing metrics \emph{downstream} from training, such as empowering individuals to achieve their creative goals. 
Instead, progress is often measured with proxy metrics that are easier to automate, such as the ability of a model to compress a dataset or generate realistic samples that imitate the training data.
For example, in music, studying composers using deep generative models ``in-the-wild'' is a nascent field of study~\cite{sturm2019machine, huang2020ai}, and not directly incorporated into the ML research process itself.

HCI researchers have separately focused on designing better interfaces for steering existing deep generative models.
These studies often treat the model architecture (and just as importantly, the training data) as fixed, adapting off-the-shelf pretrained models.
Further, many studies have focused on the experience of the creator, such as demonstrating that interfaces for steering generation and leading the collaboration can support users' experience during co-creation, including their control, ownership, engagement, and creativity~\cite{zhou2021interactive, louie2020noviceai, oh2018lead}. In addition to self-report, separately measuring the products of creation and the effect they have on outside audience is an equally important metric.
Within the domain of deep generative models for music, it remains untested if better steering can empower creators to better communicate feelings or create compositions that sound more musical, as judged by listeners.

\begin{figure*}[t]
    \centering
    \includegraphics[width=\textwidth]{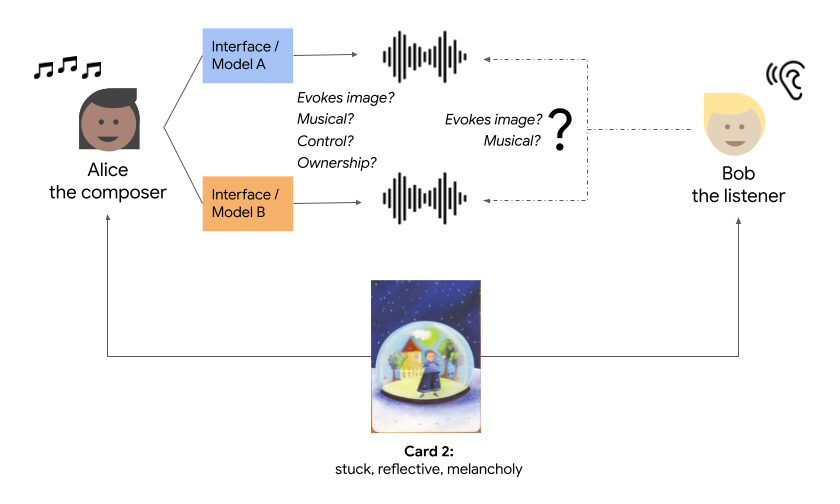}
    \caption{In the Expressive Communication framework, Alice is tasked with composing a piece of music that expresses the imagery and words of a card.  She creates two pieces of music, one for each of the different systems she is comparing, which can differ on their interfaces or their models. After completing her compositions, she provides self-report answers about her experience and the compositions made with both systems (e.g., how well does the music evoke the imagery and words? how musically coherent is your composition? how much control did you feel using this system?). Afterwards, Bob is provided the same card and listens to both music compositions made by Alice. Bob answers questions that have him compares these compositions (e.g., which music better evokes the imagery and words of the card? which sounds more musical?)}
    \label{fig:expressive_communication_framework_diagram}
\end{figure*}

\section{The Expressive Communication Framework} \label{sec:framework}
As a step towards uniting these separate developments across ML and HCI, this paper proposes {\em Expressive Communication} as a common framework for evaluating progress in generative tools (including both the model and the interface).
In Expressive Communication, we not only consider the effect of a tool on the creator's subjective experience, but also how it objectively affects what they create.
Consider the music creation scenario illustrated in Figure ~\ref{fig:expressive_communication_framework_diagram}, where ``Alice the composer'' uses generative tools to create a pieces of music that express a particular feeling and image, while ``Bob the listener'' judges which music actually best evokes that feeling and image. 
Creators self-report their perceptions of the co-creation process and their final generated music, while outside listeners evaluate that music on the basis of how well they evoke the target creative ideas and their overall musical quality. 
This expression task was chosen to align with common real-world tasks in music content creation such as creating music that establishes a mood, or creating accompanying music for another type of media like a video~\cite{frid2020music}.  
Our framework enables us to situate user studies in the context of a range of \emph{downstream} creative goals, and evaluate the effectiveness of varying both the model and interface of a tool in accomplishing these \emph{downstream} metrics, i.e. composer's \emph{subjective} self-reports and listener's \emph{objective} judgements. 

By using the Expressive Communication framework, we are able to compare both different generative models, and different interfaces to those models. As a proof of principle, we compare between two generative models capable of different degrees of long-range structure and musical coherence, and also between two different interfaces capable of different degrees of steering and iterative composition. The composers in our study created 100+ musical phrases~\footnote{Listen to the music participants composed using generative tools for different imagery and words here: \url{https://storage.googleapis.com/expressive-communication/index.html}}, 
expressing the imagery and words in the set of illustrations in Figure~\ref{fig:5cards}, which were then evaluated by listeners in 1000+ head-to-head comparisons. 

Our results show that both the ML and HCI approaches (developing better pretrained models and better steering interfaces, respectively) are important and complementary ways to support composers in both communicating through music and feeling empowered in the process of co-creating with generative models. Our results also shed light on how biases in a pretrained model capabilities such as stronger coherence can make feelings such as fear more difficult to express with curation of random samples alone, and how the addition of steering interfaces can help to overcome model biases by creating samples that are less likely from the model, but more aligned with the user's expression and musical goals.

\section{Related Work}
While the ML and HCI communities have similar aspirations in building new tools for creative use, in practice their approaches and objectives differ, making it difficult to evaluate and compare their \emph{downstream} impact on creators. 
We briefly survey how the two communities evaluate progress, their respective limitations, and show how our framework addresses some of the challenges that arise, especially in the creative context.

\subsection{ML Evaluation of Generative Models}
In ML, researchers often evaluate generative models using proxy metrics that are easy to automate. The most common objective is to maximize the likelihood of the data according to the model, which is used to measure how well a model is able to fit a desired distribution. In different contexts, this can also be interpreted as the ability of the model to maximally compress the data with the smallest reconstruction error~\cite{alemi2018information}. However, not only is likelihood an unreliable metric for comparing different model types, it is also not a measure indicative of sample quality~\cite{theis2016note}. Yet, most sequence models are trained to optimize for this metric. 
As a proxy to evaluating sample quality, researchers in image generation often use metrics based on classifiers trained on ImageNet~\cite{deng2009imagenet}, such as the Inception score~\cite{salimans2016improved} and the Fréchet Inception Distance (FID)~\cite{heusel2017gans}, to measure how well generated objects bear the features learned from classifying the objects. 
In music generation, it is common to compute simple musically informed features to measure and compare if generative systems are able to produce samples with desirable features~\cite{yang2020evaluation}. However, these metrics are only a proxy to human evaluations, and furthermore do not capture how useful the models will be \emph{downstream} to creators.

Perhaps most related to this work is the framing of Reinforcement Learning (RL) tasks~\cite{sutton2018reinforcement}. In RL, generative models are often used by agents to understand the environment and inform their actions~\cite{ha2018world}. While most of this work is focused on downstream tasks such as teaching artificial agents to play video games~\cite{Mnih2015HumanlevelCT}, an emerging thread of research is exploring ways of evaluating and optimizing for Human-AI collaboration~\cite{carroll2019utility}. Studies have show human feedback and interaction can be used to optimize generative models for collaboration on tasks such as classification or assistive game playing~\cite{reddy2021pragmatic, jaques2020learning}. This work examines a new approach towards evaluating and optimizing for Human-AI collaboration in the domain of creative expression.




\subsection{HCI Evaluation of Interactive Systems and Interfaces}

In HCI, many interactive systems and interface techniques have been explored to support human-AI creation with a generative system across domains such as drawing~\cite{oh2018lead, fan2019collabdraw}, creative writing~\cite{gero2019metaphoria}, and design ideation~\cite{koch2019may}. To evaluate these systems and techniques, researchers will conduct user studies to understand the impact they have on users' self-report feelings, their behaviors in the process, and the quality of the resulting artifacts. 

To support the evaluation of creativity tools, significant research has focused on measuring the process and experience. For example, the 
Creativity Support Index (CSI) measures the impact on six dimensions such as exploration, expressiveness, immersion, enjoyment, results worth effort, and collaboration~\cite{cherry2014quantifying}. While such research efforts have sought to provide a standard and reliable measure of comparison for a tool, psychometric indexes like CSI rely on users' subjective self-reports. Ultimately, instead of reflecting a creator's {\em subjective} perspective on the impact of the tool, we argue that evaluation frameworks that seek to understand if tools support creators effectiveness need equal emphasis for evaluating the final creative outputs.

In this vein, several methods have been employed to measure the products of creation in other domains. Within design and ideation, measures like ratings from expert judges or the quantity of products within a creative ideation have been employed~\cite{siangliulue2016ideahound}. Within interpretibility of generative models, novel tasks have been proposed for the predefined goal of interactively reconstructing a given target artifact~\cite{ross2021evaluating}, allowing researchers to compare which models enabled users to complete the task better. Nonetheless, in the context of expressive creative practices like music, such objectives have been harder to define. As such, many user studies within the music domain are situated in more open-ended creative tasks~\cite{louie2020noviceai,huang2016chordripple}. With our work, we seek to provide a comprehensive evaluation using \emph{downstream} metrics such as composer's \emph{subjective} self-reports and listeners \emph{objective} listener judgements of an expressive communication task.  In doing so, our work provides the necessary flexibility inherent in expressive creativity while also providing more objective measures upon which to evaluate the generated musical outputs.

\section{Generative Models and Steerable Interfaces}

\subsection{Steering Through Tree Search and Semantic Filtering}

\begin{figure*}[t]
    \centering
    \includegraphics[width=\textwidth]{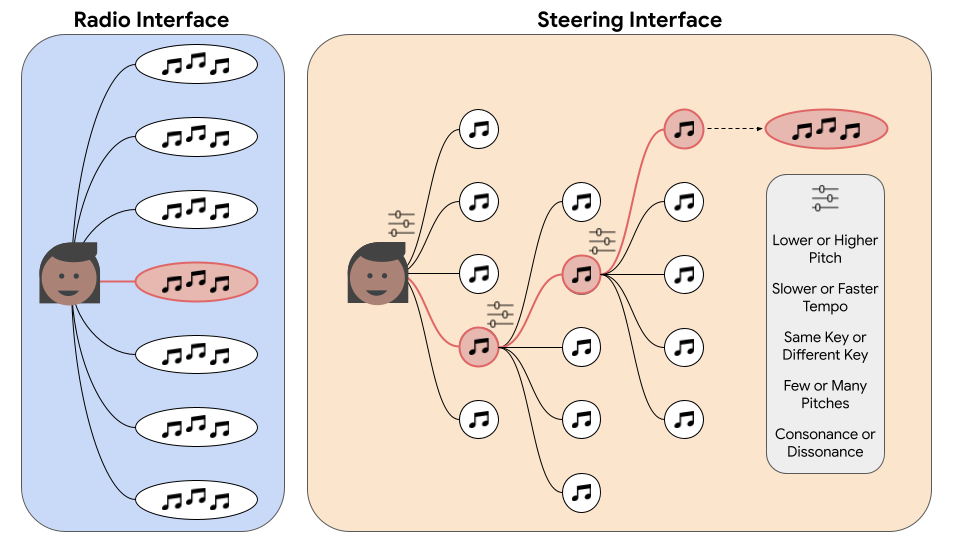}
    \caption{In our experiments, we compared two interfaces for creating music with a generative model: the Radio Interface and the Steering Interface. A composer using the Radio Interface requests randomly generated options of the full-length phrase of music, and selects the one best matches her goals. A composer using the Steering Interface will explore and select three total generated chunks for the start, middle, and end of their musical phrase. For exploring any chunk, they can request a randomly generated set, or they can choose to constrain the generation along semantic dimensions (e.g., lower or higher pitch; consonance or dissonance). After selecting any chunk, subsequent chunks are generated as a continuation of the previous chunks, in an autoregressive fashion.}
    \label{fig:interface_comparison_diagram}
\end{figure*}

Today, a common approach when interacting with a generative model is to generate a large quantity of musical samples, and curate through them to find desired results~\cite{huang2020ai}.  As a baseline approach, musicians can manually curate by listening to many alternatives, before selecting generated options which are most interesting or ``catchy'' or which might fit well for their musical goals. 
In our experiments, we implemented this baseline interface and refer to it as the {\em Radio Interface} (see Figure~\ref{fig:interface_comparison_diagram}, left), inspired by how a user chooses to listen to music from one station out of many stations on a radio~\cite{listentotransformer}.  For the Music Communication task, creators request ten randomly generated phrases that are all approximately 15 seconds and can select the option that best matches the ideas and imagery of the card. 

Recent developments in HCI research have sought to provide users (e.g., novice composers) more control and agency when creating with generative models. For example, researchers have developed steering tools for a deep generative AI models~\cite{louie2020noviceai} and showed that two core capabilities can be helpful: (1) generating music chunk-by-chunk and (2) having controls to constrain the generation along semantically-meaningful dimensions. 
Thus, we used these design principles when designing the steering tools for the autoregressive Music Transformer model.

We present a diagram of the user interaction flow for using the Steering Interface on the right-side of Figure~\ref{fig:interface_comparison_diagram}. The steering tools support listening to generated music one chunk at a time before moving onward. To do this, the autoregressive output is split into three chunks (5 seconds each), and users can choose one of the generated options for the current chunk, before seeing the next chunk which is a continuation of all subsequent chunks. Additionally, the steering tools provide users with several semantic dimensions for controlling the generation. For example, a user can select from several dropdowns on whether they want ``lower or higher pitch'', ``slower or faster tempo'', etc. These dimensions were designed to align with how novice composers think of meaningful dimensions of variation when expressing more abstract ideas and concepts through music. 
To control the generation, we use a rejection sampling approach to filter generated output along different semantic dimensions. 

For our implementation of chunk-by-chunk tree search, we used the autoregressive models to pre-generate a forest of trees with 100 parent seeds for the first chunk, 100 child continuations in the second chunk, and 100 child continuations in the 3rd chunk. This makes for 1 million possible 15 second full-length phrases that can be sampled from the model. The tool provides 10 generated options in the 1st chunk to mirror the diversity of parent nodes in the Radio, whereas the tool provides 5 generated options in the second and third chunk.

For our implementation of semantic filtering, we defined several heuristic functions that can characterize some musical feature of a chunk through a numerical value. These functions fall under two categories: absolute musical features which use the absolute numerical value (i.e., slow vs fast); and relative musical features describing a chunks musical attribute relative to its parent (i.e., make the second chunk ``slower or faster'' than the first chunk).

\begin{figure*}[t]
    \centering
    \includegraphics[trim=0 0 1cm 0, clip,width=\textwidth]{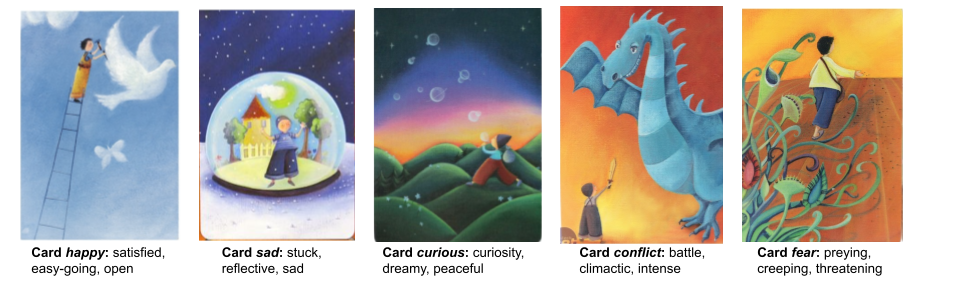}
    \caption{We selected 5 image cards that came from the board game Dixit~\cite{wiki:dixit}. They were selected to cover different parts of the valence and arousal dimensions of emotion. To ensure that composers and listeners had a common interpretation of the card, we attach three keywords to each card.}
    \label{fig:5cards}
\end{figure*}

\subsection{More Capable and Expressive Generative Models} \label{sec:models}
Within the field of generative ML, researchers have sought to develop models that are capable of more expressive timing and  dynamics~\cite{oore2020time} to models that are capable of building upon and developing expressive themes and motifs through long-range coherence~\cite{huang2018music}.
For our studies, we chose two trained autoregressive generative models capable of generating polyphonic piano music:

\begin{itemize}
    \item PerformanceRNN~\cite{oore2020time} is the less expressive model. It is trained on a smaller and more dramatic piano performance dataset (MAESTRO)~\cite{hawthorne2018enabling}, and has less capacity in the model architecture to model the distribution of music.
    
    \item Music Transformer~\cite{huang2018music} is the more expressive model. It is trained on a larger and more melodic piano performance dataset (YouTube)~\cite{youtube}, and has larger capacity to model the distribution of music.
\end{itemize}

\section{Evaluation Study}
We are broadly interested in how the Expressive Communication evaluation framework can measure the progress in ML and HCI efforts to support creative empowerment with computer generated music.

Thus we ask the following research questions: \textbf{RQ1} How are the final compositions created using (a) a more expressive generative model and (b) more steerable interface perceived by outside listeners (e.g., in evoking the desired feeling, sounding more musical); and \textbf{RQ2} How do composers feel more empowered (e.g., achieving creative goals, ownership, efficacy) when using a tool that (a) is powered by a more expressive generative model, or (b) that has more steerable interfaces?

\subsubsection{Experimental Conditions}
In this within-subjects study, our participants created music compositions in two experiments: an {\em interface comparison} between two interfaces for exerting influence on the final generated output, and a {\em model comparison} between two pretrained generative models.  For the interface comparison, participants created two pieces of music with two systems that differed in their interfaces (as illustrated in Figure~\ref{fig:interface_comparison_diagram}): a baseline {\em Radio Interface} from which they select a best match from a set of randomly generated full-length phrases; and a {\em Steering Interface} from which creators have the ability to iterate and perfect smaller sections of the phrase before moving onto selecting new ones. These two interfaces were used to steer the generated outputs of the MusicTransformer model. For the model comparison, participants created two more musical pieces with two systems that differed on their pretrained generative models: {\em PerformanceRNN} which is algorithmically more simple, less expressive and less able to elaborate a music theme; and {\em MusicTransformer} which is the more expressive model and has algorithmic capabilities that can model longer-range themes and structures in the music. To interact with these models, participants used the Radio Interface to curate and find best generated sample that matched the ideas of a card.

In our experimental setup, we counterbalanced both the ordering of the comparisons (e.g., whether they were to first use the two different interfaces, or the two different models), and the ordering within each comparison (i.e., for the interface comparison, whether they would first use the Chunks or Radio Interface). For each system comparison (interface comparison, model comparison), we assigned each participant a randomly selected card from the set of five cards; see Figure~\ref{fig:5cards}. This ensured we controlled for participants having the same expressive communication goal when comparing within a comparison.

\subsubsection{Composer Study Method}
The 26 participants who completed the study included 12 females and 14 males, ages 24 - 60 ($\mu=37$). 
Users were recruited through mailing lists at our institution and came from a variety of professional backgrounds (e.g., designer, administrator, technical writer, engineer). Each received a \$50 gift credit for their time.

The composer sessions were conducted remotely with participants over a video meeting where they shared their screen. Each user was given an overview of the goals of the study and the expressive communication game (10 minutes). They completed a guided tutorial of the systems they would be using in the first comparison (15 minutes). In the first comparison, participants were assigned a card and were asked to compose music that reflected the imagery and words of the card using each of the systems being compared for the comparison (10 minutes per system). Users were observed while composing using a think-aloud procedure. Finally, they answered a post-comparison questionnaire and completed a semi-structured interview comparing their experiences (10 minutes). This procedure was repeated for the second comparison, including being guided through a new tutorial, composing two compositions using the two systems to express a single card, and a post-comparison questionnaire and interview (40 minutes).  At the end, they answered several questions about their overall experience composing in this Expressive Communication game (10 minutes).

To test the difference between composers ratings, we conducted a two-sample paired t-test, with the null hypothesis that the mean of their differences is zero.   

\begin{figure*}[]
    \centering
     \begin{subfigure}[b]{0.48\textwidth}
         \centering
         \includegraphics[width=\textwidth]{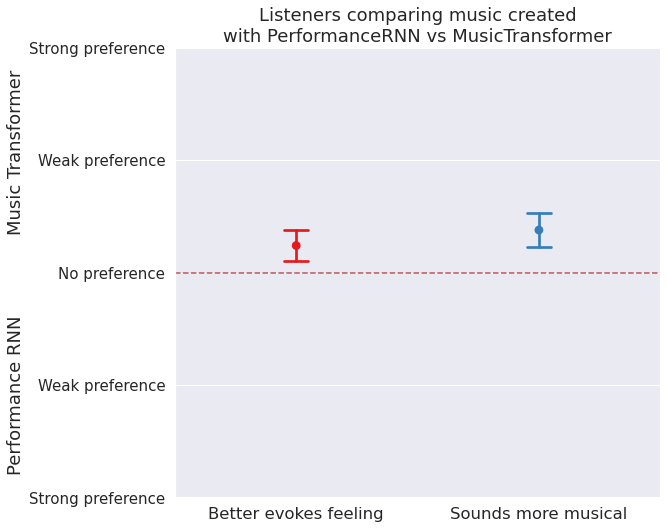}
         \caption{Listeners' ratings comparing between models.}
         \label{fig:listener_modelcomparison}
     \end{subfigure}
     \hfill
     \begin{subfigure}[b]{0.48\textwidth}
         \centering
         \includegraphics[width=\textwidth]{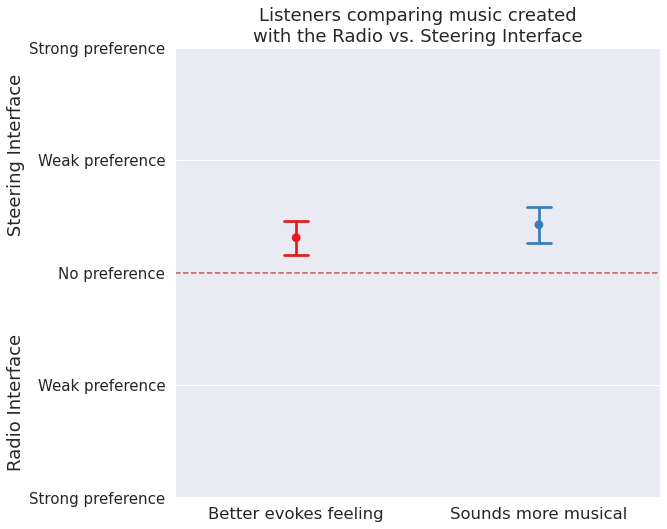}
         \caption{Listeners' ratings comparing between interfaces.}
         \label{fig:listener_interfacecomparison}
     \end{subfigure}

     \hfill
        \caption{Listeners compared compositions created for different interfaces and different generative models, along the dimensions of evoking the feelings of the card, and sounding more musical.}
        \label{fig:listener_quant}
\end{figure*}

\subsubsection{Listener Study Method}
We recruited 20 unique listeners for the listener study from an online crowd work platform. They made head-to-head comparisons for each of the pairs of musical samples created by a composer expressing the imagery and words of a particular card. After finishing our composer studies, pairs of music compositions were made for 51 comparisons (26 interface comparisons and 25 model comparisons). In total, our listeners provided 1020 head-to-head ratings comparing the pairs of music compositions.    

We asked listeners to compare the two phrases of music which were created by the same composer, for the same card. We chose this to control for the variations in composers interpretations of a given card, and to control for variations in how easy it might be to express one card vs. the other.  The ordering of the music options were randomized to prevent an association with ordering and experimental conditions.

Listeners were asked "Which one of these musical excerpts most evokes the feelings of the words and imagery on the card?" and "Which one sounded more musical" and answered on a 5 point balanced scale (``Strong preference for option 1'', ``Weak preference for option 1'', ``No preference'', ``Weak preference for option2'', and ``Strong preference for option2''). In preparation for conducting our analysis, we converted this scale to a numerical scale from $[-2, 2]$. For for the model comparison, positive values corresponded to a preference for MusicTransformer; for the interface comparison, positive values correspond to a preference for the Steering Interface. 



\section{Quantitative Results}

\subsection{Listener Study}


\subsubsection{Interface Comparison}
The listener results for the interface comparison is shown in Figure~\ref{fig:listener_interfacecomparison}. Listeners on average felt compositions made with the \textbf{Steering Interface} better evoked the feelings of the card over compositions made with the \textbf{Radio Interface}, where the difference was statistically significant ($\mu=0.31$, $t=4.09$, $p<0.0001$).  Additionally, listeners felt compositions created with the \textbf{Steering Interface} sounded more musical than those made with the \textbf{Radio Interface} ($\mu=-0.5846$, $t=3.472$, $p<0.001$). 

\subsubsection{Model Comparison}
The listener results for the model comparison is shown in Figure~\ref{fig:listener_modelcomparison}. Listeners felt compositions made with \textbf{MusicTransformer} better evoked the feelings of the card over compositions made with \textbf{PerformanceRNN}, where the difference was statistically significant ($\mu=.24$, $t=3.318$, $p<0.001$). In addition, listeners rated compositions made with MusicTransformer to sound more musical than those made with PerformanceRNN ($\mu=0.378$, $t=4.74$, $p<0.00001$).   

\begin{figure*}[t]
    \centering
    \includegraphics[width=\textwidth]{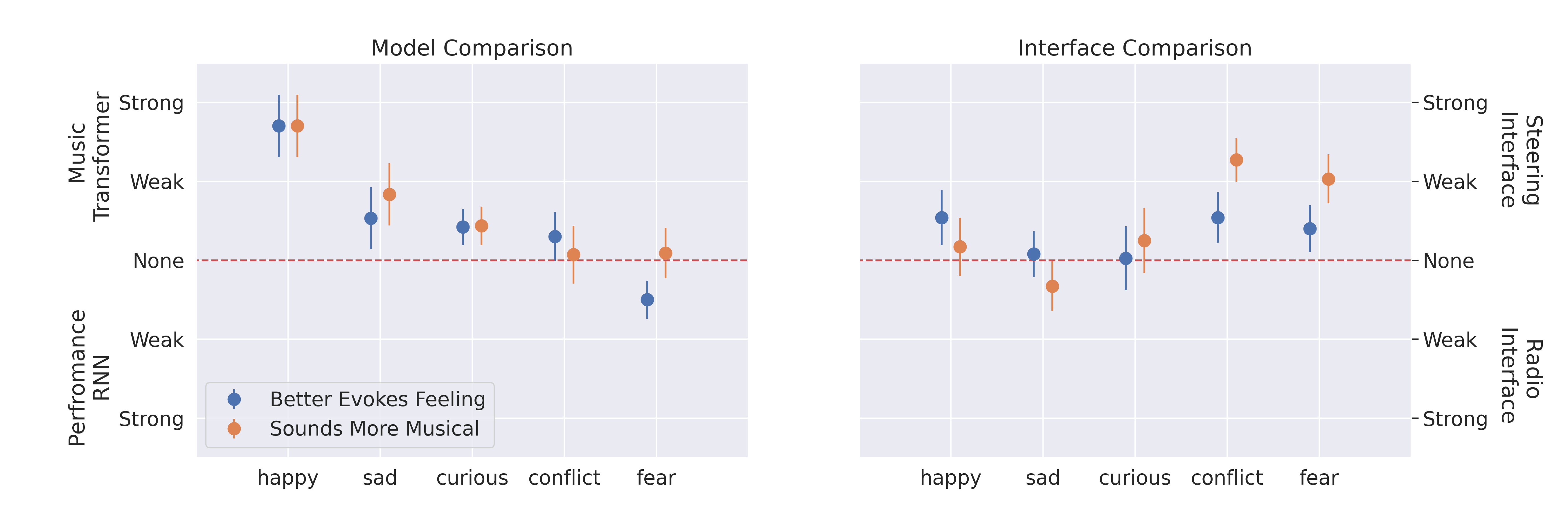}
    \caption{Listener ratings, comparing which model and interface best evoke the target feeling and/or sound more musical. For graphing, we use simplified descriptions to denote the feeling of each card. Comparing the model ratings by card (left) exposes bias in the pretrained models, where curated random samples from MusicTransformer clearly better evoke the feelings of happy, sad, conflict, and curious, while PerformanceRNN samples better evoke fear. Interestingly, these communication preferences are strongly correlated with sounding more musical for happy, sad, and curious, but there is no musical preference for conflict and fear. This exposes that the pretrained MusicTransformer has a model bias towards coherent musical output which is more aligned with straightforward feelings such as happy or sad. Further, one way of evoking conflict and fear is with musically incoherent pieces, as Performance RNN is biased to output. Comparing interface ratings across cards reveals that music created with the steering interfaces do no better than curated random samples on evoking feelings like sad and curious.}
    \label{fig:card_compare}
\end{figure*}

\subsection{Composer Study}

\begin{figure*}[t]
     \centering
     \begin{subfigure}[b]{0.48\textwidth}
         \centering
         \includegraphics[width=\textwidth]{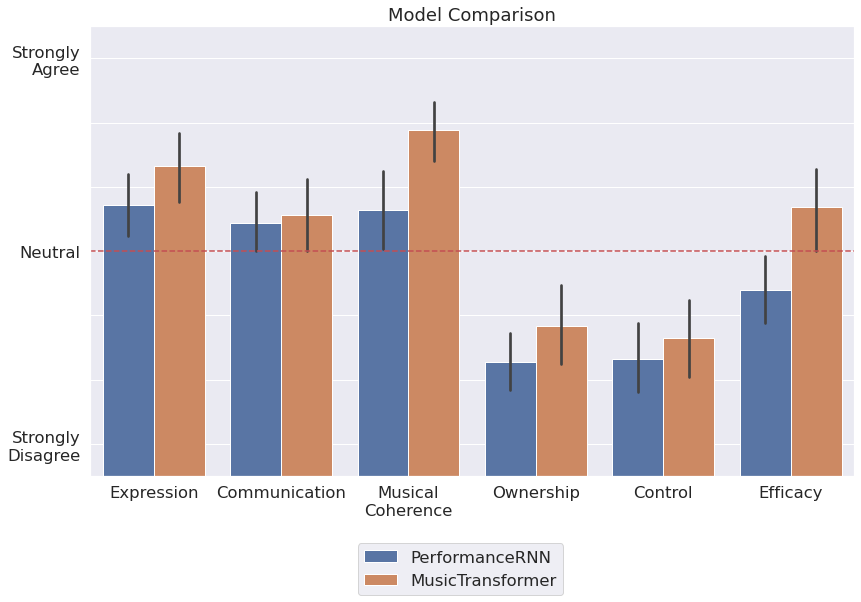}
         \caption{Composers' ratings comparing between different models.}
         \label{fig:composer_modelcomparison}
     \end{subfigure}
     \hfill
     \begin{subfigure}[b]{0.48\textwidth}
         \centering
         \includegraphics[width=\textwidth]{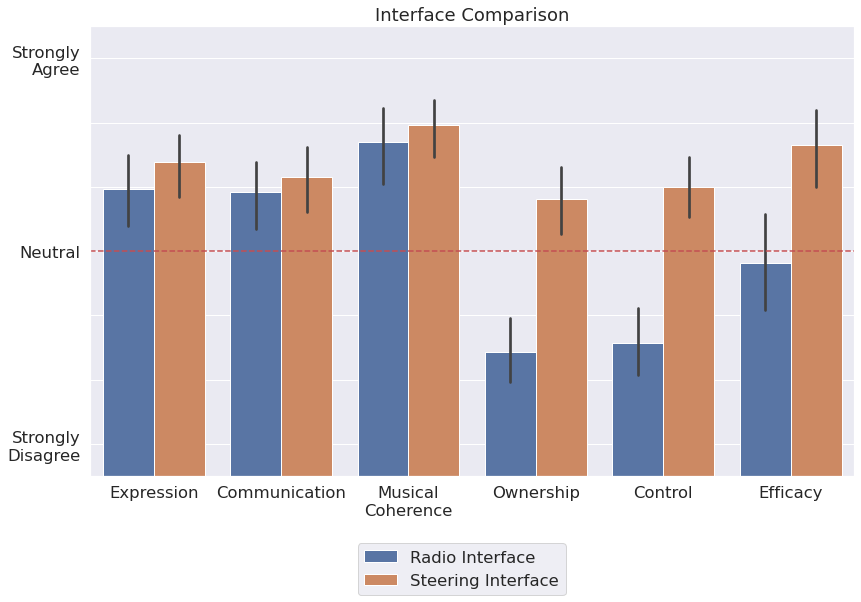}
         \caption{Composers' ratings comparing between different interfaces.}
         \label{fig:composer_interfacecomparison}
     \end{subfigure}
     \hfill
        \caption{Composers answered questions about on the final compositions made and about their experiences using the different interfaces, and different generative models.}
        \label{fig:composer_quant}
\end{figure*}

\subsubsection{Interface Comparison} The composer self-report ratings for the interface comparison is shown in Figure~\ref{fig:composer_interfacecomparison}. No significant difference was found between the final music created with either interface for \textbf{expressing} the musical qualities of the card ($\mu_1=5.38$, $\mu_2=4.96$, $t=0.98$, $p=0.33$), their confidence in the music \textbf{communicating} the ideas of the card for a listener to guess ($\mu_1=5.15$, $\mu_2=4.92$, $t=0.58$, $p=0.56$), or \textbf{musical coherence} ($\mu_1=5.69$, $\mu_2=5.96$, $t=0.96$, $p=0.35$).

Participants did feel a greater sense of \textbf{ownership} of the composition created using the Steering Interface over the one made with the Radio Interface ($\mu_1=4.8$, $\mu_2=2.42$, $t=7.7$, $p<1\mathrm{e}{-7}$), and felt they had more \textbf{control} using the Steering Interface as compared to Radio Interface ($\mu_1=5.0$, $\mu_2=2.58$, $t=6.6$, $p<1\mathrm{e}{-6}$). They also had greater \textbf{efficacy} in finding multiple solutions to achieve their goals using the Steering Interface ($\mu=4.68$) as compared to the Radio Interface ($\mu=3.4$), where the difference was statistically significant ($t=4.13$, $p<0.001$).


\subsubsection{Model Comparison} The composer self-report ratings for the model comparison is shown in Figure~\ref{fig:composer_modelcomparison}. During the study, participants found the compositions made with MusicTransformer to have more \textbf{musical coherence} ($\mu=5.88$) as compared to ones made with PerformanceRNN ($\mu=4.64$) where a paired t-test found the difference to be significant ($t=3.3$, $p<0.003$). We did not find a significant difference between the compositions made with two models on \textbf{expressing} the musical qualities of the card ($\mu_1=5.32$, $\mu_2=4.72$, $t=1.78$, $p=0.087$), or their confidence in the music \textbf{communicating} the ideas of the card for a listener to guess ($\mu_1=4.56$, $\mu_2=4.44$, $t=0.35$, $p=0.72$).  

Participants felt greater \textbf{ownership} of the composition created using MusicTransformer over the one made with PerformanceRNN ($\mu_1=2.84$, $\mu_2=2.28$, $t=3.22$, $p<0.004$), and felt they had more \textbf{control} using MusicTransformer ($\mu=2.64$) as compared to PerformanceRNN ($\mu=2.32$). They also had greater \textbf{efficacy} in finding multiple solutions to achieve their goals using MusicTransformer ($\mu=4.68$) as compared to PerformanceRNN ($\mu=3.4$), where the difference was statistically significant ($t=5.02$, $p<.0001$).

\subsection{Discussion of Composer and Listener Study Results}
In sum, providing composers with a more steerable interfaces to the generation helped them feel more ownership, control, and efficacy in finding multiple solutions.  
Moreover, providing composers with a better model (while keeping the baseline Radio Interface constant) resulted in a small but statistically significant difference in ownership and control, which composers attributed to the model generating more viable options to choose from.  Given that better interfaces and better models matter to these measures of empowering users, we emphasize that providing a more steerable interface led to a bigger increase in ownership, control, and efficacy. 

For creating a composition that better evokes the feeling of the imagery and words {\em when a listener hears it}, using a system with a more steerable interface, or a more expressive model, {\em both make a difference}.  However, {\em from composers' self report}, no overall difference in expressiveness or communication was found between music created with one system versus the other, across interfaces and models.  

For creating a composition that sounds more musical {\em to listeners ears}, using a system with a more steerable interface, or a more expressive model, again {\em both make a difference}. From {\em the perspective of composers}, compositions made with the Music Transformer, a model capable of maintaining long-term structure in the music, also makes a difference for musical coherence; however, no difference was found in composer-perceived musical coherence between compositions made with the different interfaces.




\section{Core Takeaways}

\subsection{Developments in Models and Interfaces are Complimentary for Creative Empowerment and Effectiveness}

We used the Expressive Communication Framework to study how using tools powered by more expressive autoregressive models, or better steering interfaces for these models can make a difference.
Our results show that both better steering interfaces and more expressive models make a difference in composer’s feelings of empowerment (i.e., control, ownership, efficacy) and their effectiveness in creating music that evokes the intended feelings and that sounds more musical to outside listeners.

\subsubsection{Better Steering Interfaces Empower Composers} Better Steering Interfaces empowered composers by making a positive impact in their feelings of control, ownership and efficacy; see the measurable quantitative difference in Figure~\ref{fig:composer_interfacecomparison}.
Participants using the steering interface felt they \textit{``could more piece it together''} through the ability to make choices for each chunk, and said they could \textit{``compose a flow, a narrative''} and \textit{``think about the direction it takes''} (P7). Through allowing the composer to be more involved in incrementally building up the piece, composers could see how ``different elements show up in the different chunks that I chose'' (P12).  This extra involvement and work to match the music helped composers feel more ownership: ``because I put that extra effort and put more thoughts into it, I liked it better'' (P22). 

In comparison, composing by curating random outputs provided very little control. 
Since composers had little ability to provide input beyond making a final selection, some participants described their role more as a \textit{``listener and evaluator''}, rather than as composer or collaborator.  This mode (or lack thereof) of interaction coincided with a loss of ownership when using the Radio interface. As one participant said, \textit{``since I showed my preference only a little bit in choosing from the different examples, I don't feel it was mine''} (P4).

From our quantitative results, creators felt they had increased efficacy in finding multiple music phrases that matched the ideas and mood of the card when using the Steering Interface. For composers who did find multiple solutions, a common strategy was to focus on finding a first chunk that captured the general mood and tempo, through a combination of semantically constraining the generation and curating to find a match. From there, composers felt confident that they had set a good initial direction, and often noticed that many generated options in subsequent chunks continued the express the same feeling established by the first chunk.

Interestingly, since creators could find multiple options in the subsequent chunks that matched the ideas and mood, they had greater flexibility to control the musicality of the piece and use their musical preferences to guide the final output.  For example, a participant mentioned that \textit{``because I could find many phrases that match, I had room to think about how polished and consistent the phrases and chunks were''} (P3). As another way of focusing on musical coherence, many participants when composing the last chunk felt it was important to provide an ending that resolved the musical phrase, and so focused selecting a generated option that satisfied this musical goal.  For example, one participant who was expressing the feeling for ``satisfied'' said \textit{``I really like this third chunk, which ended on a proud note. It is very consonant and does resolve''} (P15). 


\subsubsection{Coherent and Expressive Models Help Evoke Imagery and Ideas}

When reflecting on the difference between models, composers described MusicTransformer as capable of evoking a coherent musical idea. For example, a participant said MusicTransformer communicated a \textit{``single thread''} which \textit{``repeats and builds upon a small theme,''} ultimately helping to evoke a clear image and mood (P5). 
In contrast, composers commented that the musical phrases generated by PerformanceRNN were disjoint between sections, were less coherent, and \textit{``don't know what they want to be''} (P12). Another participant elaborated that the music samples from PerformanceRNN would \textit{``often change the tempo or tonality too fast, and there wasn't a consistent feeling in it''}, which made it difficult to find samples that \textit{``expressed something that was clearly evocative''} (P8). As we found in our quantitative results, participants curating random samples from PerformanceRNN had lower efficacy in finding musical samples that matched the creative ideas they were trying to express. After asking composers to reflect on their experience using PerformanceRNN, they felt that since ``there were fewer [options] that I was happy with, I felt like I did have less control and fewer solutions'' (P21). Some participants elaborated that the lack of a consistent theme in the generated music made it difficult to find many candidates since \textit{``in each of them there's always something, like an interjection, that does not quite match the card''} (P9). 

Our results also highlight how models that generate music that are expressive and relatable also helped composers with the expressive communication task. Participants felt that MusicTransformer tended to generate more modern music or pop-songs which were easier for them relate to.  For example, composers sometimes would connect a certain music to a particular cultural reference, such as reminders to specific songs they have heard (e.g., ``it sounds very similar to the song `Silent Night', so I already imagine snow and winter'') and types of music they have heard used in specific contexts (e.g., ``this is music that plays at the beginning of a Japanese anime drama''; ``it reminds me of the music that plays during a boss battle in a video game''). They could use these connections to other contexts to guide their decision on whether this music expressed the matching imagery or words of the cards.  In contrast, participants noticed that the PerformanceRNN model would generate music that associated with \textit{``classical music''} or a \textit{``20th century modern style} (P8). While the generated music was complex and sophisticated, it did not express something that was clearly evocative. These findings shed light on the importance of the cultural-relatedness and expressive moods of the pretrained models from which co-creation tools are built upon.

\subsection{Steering can Help Overcome Model Biases}

\subsubsection{Biases in Pretrained Models Help to Better Evoke Some Feelings Over Others} Comparing model preferences by card (Figure~\ref{fig:card_compare}, left) exposes a bias in pretrained models, where curated random samples from MusicTransformer evoke the feelings of happy, sad, curious, and conflict, while PerformanceRNN samples better evoke fear. Interestingly, effectively evoking the feeling is strongly correlated with sounding more musical, revealing that pretrained MusicTransformer has a bias towards coherent musical output which is more aligned with straightforward feeling such as happy or sad. 

From the perspective of composers, several felt that MusicTransformer was especially useful for expressing imagery and ideas from some cards, such as dreamy, curiosity, sad, easy going, and satisfied. 
However, several participants felt that MusicTransformer's generated outputs could be \textit{``too steady''}, which they felt \textit{``worked well for walking or peaceful but less so for something with a climax''} (P12). In contrast, participants using PerformanceRNN took advantage of the bias in its generated outputs towards feelings like fear using \textit{``rhythms that are really not steady... with short-long-short-long patterns that interfere with each other''} which they felt made this sample \textit{``the creepiest''} (P14). 



\subsubsection{Steering Interfaces are More Helpful When Expressing Feelings that are Misaligned with Model Biases}
Comparing interface ratings across cards (Figure~\ref{fig:card_compare}, right) shows that when the goal is to express feelings like sad and curious, music created with the steering interfaces do no better than curated random samples. 
This reveals that when random samples from an expressive model easily aligns with a goal, it can work just as well as steering to express those same feelings. This finding is corroborated with evidence from our composer study, where composers whose goal was to express straightforward feelings felt they could achieve their same goal just as well, but in a different way, when curating random samples generated from MusicTransformer rather than steering. In these cases, they were open-minded to \textit{``ceding control and seeing where it takes me''} by letting the model generate music that surprised them and was different than their initial expectations (P1). For example, another participant said \textit{``I was surprised by the random composition, taking it into a musical direction that I would not have otherwise gone... this tool makes me go beyond my first idea, giving ideas that help one express an idea in a different way''} (P15).

On the contrary, when the goal is to express feelings like conflict and fear that are less likely to be represented in the random samples, steering interfaces made a difference in creating music that better evokes the feeling (see Figure~\ref{fig:card_compare}, right). Observations during the talk-aloud composers sessions help to provide additional evidence of how steering can help to express these feelings. As an example of a steering strategy to express the feelings of fear, a composer selected a first chunk that is \textit{``slower and has less energy to signal something impending that hasn't happened yet''}, added more feelings of ominous by setting the semantic parameters to \textit{``slower, lower, and different key''}, and steered the third chunk to be faster and much lower in the end to \textit{``build tension at the end to signal that the character is being attacked by the killer plants''} (P3). 

Interestingly, the compositions made through steering interfaces sound more musical too when the goal is to express feelings like fear and conflict. We observed in the composer studies that participants using curation of random samples would try to convey “fear” with music that would keep the listener uncomfortable or not knowing what would come next. For example composers picked up on \textit{“short long note patterns”} and \textit{“sudden pauses and tempo shifts”} to distinguish from the other samples; thus, they would sacrifice coherence in this context to evoke this emotion.
In contrast, with the steering interface, they would find solutions that satisfied both the “fear feeling”, through interesting dynamics and developments across the chunks, while also retaining a musically sophisticated sound.  As one participant described, \textit{``The [music created with] chunks could be more coherent when compared to the radio version... the radio version has the right elements, but because of that, they aren't glued together''} (P3).

\section{Discussion and Future Work}





\subsubsection{Benchmarking Across Studies with Objective Listener-Centric Metrics}
The Expressive Communication framework can be commonly applied across more interactive tools and models as other researchers develop them as a useful measure for how well they do. 
We anticipate that the Expressive Communication framework can promote better cross comparisons across various music co-creation studies.  For example, researchers and developers of new intelligent user interfaces for music co-creation can task participants with generating music pieces according to our Expressive Communication framework using the imagery and words of the cards used in our studies~\ref{fig:5cards}. Then, researchers can investigate how the music created using their new model or interface can better evoke the ideas as compared to the generated musical outputs from our studies as a baseline.  


\subsubsection{Objective for Training New Collaborative Human-AI Systems}
A new frontier for ML research is explicitly training models to optimize downstream Human-AI collaborative tasks~\cite{carroll2019utility, jaques2020learning, reddy2021pragmatic}. 
As human interaction is expensive and does not scale well to training, these works use limited human interactions and evaluations to learn approximations of human behavior and preferences that can be used for automated training~\cite{christiano2017deep, stiennon2020learning}.
The Expressive Communication framework could be adapted to these approaches to enable human evaluation to scale to providing direct training signal for adapting models to better communicate and create more musical samples. 
This line of work often considers the interface to be fixed and optimize a model for the collaboration objective.
One future avenue to possibly bring together ML and HCI research would be to use HCI insights to parameterize a space of possible interfaces and use agents (with policies distilled from imitating humans) as ``proxy users'', enabling automated optimization of a new user interface that best allows a proxy user to satisfy the approximate human evaluation objective.
Such a system would enable ML and HCI researchers to better utilize limited human interaction and evaluation to bootstrap new systems where the models and interfaces are co-optimized for better Human-AI collaboration.
While this line of thinking is speculative, this work represents a first step towards that direction, demonstrating the value of evaluating generative tools with a framework that features both human demonstration and evaluation towards a common objective.



\bibliographystyle{ACM-Reference-Format}
\bibliography{main}

\appendix

\end{document}